\documentclass{birkjour}
 \theoremstyle{definition}
 
 \theoremstyle{remark}
 
 \newtheorem{ex}{Example}
 \numberwithin{equation}{section}
 \usepackage{indentfirst}

\title[On presymplectic structures for massless higher-spin fields]
{On presymplectic structures for massless \\higher-spin fields}
\author{Alexey A. Sharapov}
\address{Physics Faculty, Tomsk State University, Tomsk 634050, Russia}
\email{sharapov@phys.tsu.ru}

\begin{document}
\maketitle
\begin{abstract}
 A natural presymplectic structure for non-Lag\-ran\-gian equations of motion governing the dynamics of free higher-spin fields in four-dimensional anti-de Sitter space is proposed. This presymplectic  structure is then used to the derivation of the conserved currents associated with the relativistic invariance  and to the construction of local functionals of fields that are gauge invariant on shell.
\end{abstract}

\section{Introduction}

The higher-spin (HS) gauge theories are the general-covariant
field-theoretical models involving massless fields of spin $s>2$.
The nonlinear equations of motion for massless HS fields were
proposed by Vasiliev in \cite{V}. They exhibit some rather unusual
properties compared to the low spins:
\begin{itemize}
  \item The equations cannot be consistently perturbed about flat space-time; the most symmetrical vacuum solution is that corresponding to the (anti-)de Sitter space with nonzero cosmological constant.
  \item The dynamical content is given by an infinite spectrum of fields of increasing spins, admitting no finite truncation with fields of spin $s>2$.
  \item After exclusion of auxiliary fields, the interaction vertices and the gauge symmetry transformations involve arbitrary high space-time derivatives of dynamical fields.
\end{itemize}

The infinite number of interacting fields together with the higher
derivatives may pose some technical difficulties, but the real
challenge is the non-Lagrangian form of the Vasiliev equations. It
is the absence of a closed Lagrangian formulation which hampers our
understanding of the quantum properties of HS theories and prompts a
search for alternative quantization methods that are not rigidly
bound to the Lagrangian form of dynamics. One of such methods was
proposed in \cite{KazLS}. It is based on the concept of Lagrange
structure, which may be thought of as a strong homotopy
generalization of the Batalin-Vilkovisky antibracket. In
\cite{KLS0} and \cite{KLS2},  this quantization method was applied
to the unfolded representation of the scalar field theory and to the
Bargmann-Wigner equations for massless fields of spin $s\geq 1/2$.
Unfortunately,  this approach, while general, becomes unduly
cumbersome when applied to HS theories in unfolded representation.

In the present paper, an alternative method of quantizing unfolded
HS dynamics is developed. It exploits the notion of covariant
presymplectic structure pioneered by Crnkovi\'c \& Witten \cite{CW}
and independently by Zuckerman \cite{Z}. An extensive historical
overview of the subject and further references can be found in
\cite{Kh0}. Specifically, we show that the free HS fields on
four-dimensional anti-de Sitter space admit quite a natural
presymplectic structure which is compatible with the unfolded
representation of HS dynamics and generalizes the covariant
presymplectic structures for low-spin theories. As is well known,
every presymplectic structure gives rise to a Poisson bracket in the
space of gauge invariant functionals of fields and can be quantized,
in principle, by means of the deformation quantization technique.  There
is reason to hope that the proposed presymplectic structure admits a
consistent extension to the interacting HS fields.  If this is the
case we get a good starting point for the covariant quantization of
nonlinear HS dynamics. 

The concept of covariant presymplectic structure is found to be useful  in the study of some other aspects of HS dynamics, not directly related to the problem of quantization.  For example, every presymplectic structure is known to provide a systematic correspondence between symmetries and conservation laws. So far such a correspondence has been established indirectly just by comparing tensor parameters entering the conserved currents and the symmetry transformations of the free HS fields. Furthermore, given a presymplectic structure, it is possible to define a local functional of fields whose stationary surface includes all the solutions to the original field equations \cite{BHL, Kh}. It is hoped that functionals of this kind may find applications in the context of AdS/CFT correspondence \cite{M, GKP, W,V3}.

The plan of the paper is as follows. In Sec. \ref{PrM}, we provide
some background material on presymplectic geometry and its relation
to classical mechanics. Next, in Sec. \ref{CPhS}, we briefly discuss
the covariant phase-space approach to field theory and, by way of
illustration, derive the covariant presymplectic structures for
various fields of low spins. There,  we also define a general notion
of covariant presymplectic structure for not necessarily Lagrangian
field theories. In Sec. \ref{HSFF}, we review the unfolded
representation for the free HS fields on four-dimensional anti-de
Sitter space. The main result of the paper is presented in Sec.
\ref{CPS}, where we derive a covariant presymplectic structure for
the free HS equations in unfolded representation. In Sec. \ref{SCL},
this presymplectic structure is used to define the gauge
noninvariant conserved currents associated with the anti-de Sitter
invariance of the free HS equations. As a byproduct this proves
nontriviality of the proposed presymplectic structure. Sec. 7 is devoted to the derivation of on-shell gauge invariant functionals of HS fields. 
In the last
Sec. \ref{Sum}, we summarize our results and discuss further
perspectives.

\section{Presymplectic mechanics}\label{PrM}

Let us briefly recall some basic notions of presymplectic geometry
and geometrical mechanics.

By definition, a \textit{presymplectic manifold} is a pair
$(M,\Omega)$ consisting of a smooth manifold $M$ endowed with a
closed $2$-form $\Omega$, the \textit{presymplectic form}. Denote by
$\ker \Omega$ the space of all vector fields $V$ on $M$ satisfying
the equation
$$
i_V\Omega=0\,.
$$
The form $\Omega$ being closed,  the space $\ker \Omega$ generates
an integrable distribution on $M$. In the  case that $\ker \Omega
=0$ one speaks of the {symplectic} $2$-form $\Omega$ and the
symplectic manifold $M$.

A vector field $X$ (a function $f$) is called \textit{Hamiltonian}
if there exists a function $f$ (a vector field $X$) such that
\begin{equation}\label{xof}
i_X\Omega=df\,.
\end{equation}
In order to indicate the relationship between the Hamiltonian vector
fields and functions we will write $X_f$ for $X$ and refer to $f$ as
a Hamiltonian of the vector field $X_f$.  Note that the above
correspondence $f \leftrightarrow X_f$ is far from being one-to-one.
On the one hand we can add to $X$ any vector field from $\ker
\Omega$ without any consequence for $f$, on the other it is possible
to shift $f$ by an arbitrary constant. This motivates us to consider
two Hamiltonian fields as equivalent if they differ by an element
from $\ker \Omega$. By definition, the space $\ker \Omega$ consists
of the Hamiltonian vector fields with constant Hamiltonians. In what
follows we will often identify a Hamiltonian vector $X_f$ with its
equivalence class $X_f+\ker \Omega$.   It is easy to see that the
Hamiltonian vector fields form a subalgebra $\mathfrak{X}_\Omega(M)$
in the Lie algebra of all smooth vector fields $\mathfrak{X}(M)$.
Furthermore, $\ker\Omega$ is an ideal in $\mathfrak{X}_\Omega(M)$
and  we can define the quotient Lie algebra
$\mathfrak{X}_\Omega(M)/\ker\Omega$ of ``nontrivial Hamiltonian
vector fields''.

An elementary, yet important, fact is that the action of a Hamiltonian
vector field preserves the presymplectic structure. Indeed,
\begin{equation}\label{LXFO}
L_{X_f}\Omega =di_{X_f}\Omega+i_{X_f}d\Omega=d^2f=0\,.
\end{equation}

Another simple observation is that Hamiltonian functions form a
commutative algebra with respect to the point-wise multiplication:
If $f$ and $g$ are two Hamiltonian functions, then $X_{fg}=f\cdot
X_g+g\cdot X_f$. The space of Hamiltonian functions can be  endowed
with the Poisson bracket
\begin{equation}\label{PB}
\{f,g\}=i_{X_f}i_{X_g}\Omega \,.
\end{equation}
One can easily verify that this expression is well defined and
satisfies all the required properties: antisymmetry, bilinearity,
the Leibniz rule, and the Jacobi identity.  Denoting  the Poisson
algebra of Hamiltonian functions by $\mathcal{F}_\Omega(M)$, one can
also see that the map $f\mapsto X_f$ is actually a homomorphism of
the Lie algebras, meaning that  $X_{\{f,g\}}=[X_f,X_g]$.

From the physical viewpoint, the Hamiltonian functions represent the
physical observables. It follows from the definition (\ref{xof})
that each physical observable is invariant under the action of the
integrable distribution $\ker \Omega$, i.e.,
$$
Xf=0\qquad \forall X\in \ker \Omega \,,\quad \forall f\in \mathcal{F}_\Omega(M) \,.
$$
The vector fields of $\ker \Omega$ play thus the role of infinitesimal gauge symmetry transformations on $M$ and the integral leaves of the distribution $\ker\Omega $ should be regarded as gauge orbits. The physical phase space is then identified with the space of gauge orbits, though the latter may not be a smooth manifold in general. According to this interpretation the physical observables are those smooth functions on $M$ that are constant along the gauge orbits. Therefore, they can be viewed as functions on the physical phase space $M'=M/\ker \Omega$ and the Poisson bracket (\ref{PB}) passes through the quotient.

In order to define the time evolution of physical observables one
needs to specify a particular vector field $X$ that leaves invariant
the presymplectic structure, that is,
\begin{equation}\label{LXO}
L_X\Omega=0\,.
\end{equation}
Then  the equation of motion for an observable $f$ reads
\begin{equation}\label{Xf}
\dot f= Xf\,.
\end{equation}
Each vector field $X$ satisfying (\ref{LXO}) is called
\textit{locally Hamiltonian}. From (\ref{LXFO}) it follows  that
each Hamiltonian vector field is locally Hamiltonian. Eq.
(\ref{LXO}) is clearly equivalent to $di_X\Omega=0$, which implies
the existence of a smooth function $f$ such that
$i_X\Omega|_U=df|_U$ for any contractable open domain $U\subset M$
(the Poincar\'e Lemma). In other words, each locally Hamiltonian
vector field becomes Hamiltonian when restricted to a small
vicinity, hence the name.

The flow generated by a locally Hamiltonian vector field $X$ defines
a one-parameter group of automorphisms of the Poisson algebra
$\mathcal{F}_\Omega(M)$, provided $X$ is complete. In case $X$ is
Hamiltonian, that is, $X=X_h$ for some $h\in \mathcal{F}_\Omega(M)$,
Eq. (\ref{Xf}) assumes a more familiar form
$$
\dot f=\{f,h\}\,.
$$

Bringing the classical dynamics into the Hamiltonian form is usually considered as ``a must'' step towards quantization.

\section{Covariant phase space}\label{CPhS}

Recall that in the conventional Hamiltonian formalism the phase
space of fields is identified with the space of Cauchy data to the
field equations. This requires  a prior splitting of physical
space-time into space and time, violating thus the relativistic
invariance.  This drawback is avoided in the covariant approach to
Hamiltonian mechanics, where the phase space of fields is identified
with the space of solutions to the field equations, rather than the
Cauchy data. Under certain technical assumptions, like global
hyperbolicity of the underlying space-time manifold, these two
spaces may be viewed as equivalent. For Lagrangian equations of
motion the solution space comes equipped with a natural presymplectic structure,
making the space of gauge invariant functionals of fields into a
Poisson algebra.

 Consider, for example, the action functional
\begin{equation}\label{S}
S[\phi]=\int_V  \mathcal{L}(\phi^i, \phi_\mu^i) d^nx
\end{equation}
for a collection of bosonic fields $\phi^i(x)$. Here
$\phi^i_\mu=\partial_\mu\phi^i$ and integration is performed over a
bounded domain  $V$ in an $n$-dimensional space-time manifold with
local coordinates $x^\mu$. Varying the action, we get
\begin{equation}\label{EqMot}
\delta S=\int_V \left(\frac{\partial \mathcal{L}}{\partial \phi^i} -\partial_\mu\frac{\partial \mathcal{L}}{\partial \phi_\mu^i}\right)\delta \phi^i \wedge d^nx+\int_{\partial V}\frac{\partial \mathcal{L}}{\partial\phi^i_\mu}\delta\phi^i\wedge d^{n-1}x_\mu \,.
\end{equation}
Hereafter we use the following properties of the de Rham and variational exterior differentials:
\begin{equation}\label{dd}
\quad d^2=\delta^2=0\,, \qquad  d\delta=-\delta d\,,\qquad \partial_\mu\delta = \delta\partial_\mu\,,
\end{equation}
$$
dx^\mu\wedge dx^\nu=-dx^\nu\wedge dx^\mu\,,\quad \delta\phi^i\wedge \delta\phi^j=-\delta\phi^j\wedge\delta \phi^i\,,\quad dx^\mu\wedge \delta\phi^i=-\delta\phi^i\wedge dx^\mu\,.
$$
The bulk term in (\ref{EqMot}) defines the classical equations of motion
\begin{equation}\label{feq}
\frac{\partial \mathcal{L}}{\partial \phi^i} -\partial_\mu\frac{\partial \mathcal{L}}{\partial \phi_\mu^i}=0\,,
\end{equation}
while the boundary term gives rise to a functional $1$-form
$$
\Theta[\phi,\delta\phi] =\int_{\Sigma} \frac{\partial \mathcal{L}}{\partial\phi^i_\mu}\delta\phi^i\wedge d^{n-1}x_\mu\,,
$$
with $\Sigma$ being a Cauchy surface. To make contact with the notation of the previous section, we denote by $M$ the space of all solutions to the field equations (\ref{feq}). This will be considered as an infinite-dimensional submanifold in the space of all field configurations $\Phi$.
Some authors refer to $\Phi$ and $M$ as the spaces of \textit{all} and \textit{true} histories, respectively. We will call $M$ the \textit{dynamical shell} or just \textit{shell}. Applying the variational differential to $\Theta$ gives the functional $2$-form on $\Phi$
\begin{equation}\label{Omega}
\begin{array}{c}
\displaystyle\Omega =\delta \Theta = \int_\Sigma \delta \left( \frac{\partial \mathcal{L}}{\partial \phi^i_\mu}\right)\wedge \delta\phi^i\wedge d^{n-1}x_\mu\\[5mm]
\displaystyle =\int_\Sigma \left(\frac{\partial^2 \mathcal{L}}{\partial\phi^j\partial \phi_\mu^i}\delta\phi^j\wedge \delta\phi^i+\frac{\partial^2 \mathcal{L}}{\partial\phi_\nu^j\partial \phi_\mu^i}\delta\phi_\nu^j\wedge \delta\phi^i\right)\wedge d^{n-1}x_\mu\,.
\end{array}
\end{equation}
By construction, the $2$-form $\Omega$ is $\delta$-closed and, upon restriction to $M$, endows the solution space with a presymplectic structure. For simplicity we will denote the restriction $\Omega|_M$ by the same symbol $\Omega$. The functional $1$-form $\Theta$ is called the \textit{presymplectic potential}.

An important property of the form $\Omega$ is its on-shell independence of the Cauchy surface $\Sigma$. Let $\Phi$ be the space of fields that vanish at spatial infinity together with their derivatives and let $\Omega_\Sigma$ and $\Omega_{\Sigma'}$ denote two presymplectic forms associated with nearby space-like hyper-surfaces $\Sigma$ and $\Sigma'$.  By Stocks' theorem
$$
\Omega_\Sigma-\Omega_{\Sigma'}=\int_V d\left(\delta\frac{\partial \mathcal{L}}{\partial \phi^i_\mu}\wedge \delta\phi^i\right)\wedge d^{n-1}x_\mu\,,
$$
where $\partial V=\Sigma-\Sigma'$. Using formulas (\ref{dd}) and the identity $$dx^\nu\wedge d^{n-1}x_\mu=\delta^\nu_\mu d^nx\,,$$ we find
$$
\begin{array}{rcl}
\Omega_\Sigma-\Omega_{\Sigma'}&=&\displaystyle\int_V \left(-\delta \left(d\frac{\partial \mathcal{L}}{\partial \phi^i_\mu}\right)\wedge \delta\phi^i +\delta \left(\frac{\partial \mathcal{L}}{\partial \phi^i_\mu}\right)\wedge \delta d\phi^i\right)\wedge d^{n-1}x_\mu\\[5mm]
&=&\displaystyle\int_V \left(\delta \left(\partial_\mu\frac{\partial \mathcal{L}}{\partial \phi^i_\mu}\right)\wedge \delta\phi^i +\delta\left(\frac{\partial \mathcal{L}}{\partial \phi^i_\mu}\right)\wedge \delta \phi_\mu^i\right)\wedge d^{n}x\\[5mm]
&\approx&\displaystyle\int_V \left(\delta \left(\frac{\partial \mathcal{L}}{\partial \phi^i}\right)\wedge \delta\phi^i +\delta\left(\frac{\partial \mathcal{L}}{\partial \phi^i_\mu}\right)\wedge \delta \phi_\mu^i\right)\wedge d^{n}x\\[5mm]
&=&\displaystyle\int_V  \delta^2 \mathcal{L}\wedge d^nx=0\,.
\end{array}
$$
Here we introduced the sign ``$\approx$'' meaning the equality ``on shell''.

If there is no gauge symmetry in the theory, then the Hessian matrix
$$
\left(\frac{\partial^2 \mathcal{L}}{\partial \phi^i_0\partial\phi^j_0}\right)
$$
is nondegenerate and one can pass directly to the Hamiltonian formalism. It can be seen that the canonical symplectic structure on the phase space of fields and conjugate momenta is essentially equivalent to the presymplectic structure (\ref{Omega}). In particular, the on-shell presymplectic form $\Omega$ appears to be nondegenerate in this case and defines actually a symplectic structure.

For gauge invariant action functionals (\ref{S}) the corresponding $2$-form $\Omega$ is necessarily degenerate \cite{LW}, \cite{T}. This can be seen as follows.
For each solution $\phi\in M$, the on-shell presymplectic structure defines an antisymmetric bilinear form $\Omega[\delta\phi,\delta\phi]$ on the tangent space $T_\phi M$. Let $\delta_\zeta\phi$ be an infinitesimal gauge transformation of fields, with $\zeta$  being an arbitrary gauge parameter. Since the gauge transformations map solutions to solutions,  $\delta_\zeta \phi\in T_\phi M$. We want to show that $\Omega[\delta\phi,\delta_\zeta\phi]=0$ for all $\phi\in M$ and $\delta\phi \in T_\phi M$. By definition, the functional $\Omega[\delta\phi,\delta_\zeta\phi]$ is given by the integral over $\Sigma$ of a smooth form  depending linearly on  $\zeta$ and its derivatives.  This means that $\Omega[\delta\phi,\delta_\zeta\phi]=0$, whenever $\Sigma\cap \mathrm{supp}\, \zeta=\emptyset$. On the other hand,  we have shown that the on-shell presymplectic form does not depend on the choice of the hyper-surface $\Sigma$; and hence, $\Omega[\delta\phi,\delta_\zeta\phi]$ must vanish for \textit{any} compactly supported function $\zeta$. In view of locality, the last fact implies that $\Omega[\delta\phi,\delta_\zeta\phi]$ vanishes identically for arbitrary gauge parameter $\zeta$ and every infinitesimal gauge transformation corresponds to  a degenerate direction for the on-shell presymplectic form $\Omega$.

As usual one can factor out the presymplectic manifold $M$ by the action of gauge transformations and obtain thus the \textrm{physical phase} space $M'$. The latter is equipped with the symplectic structure $p_\ast (\Omega)$ given by the pull-back of the on-shell $2$-form $\Omega$ with respect to the canonical projection $p: M\rightarrow M'$. The physical observables, i.e., functions on $M'$, can then be identified with the gauge invariant function(al)s on $\Phi$; in so doing, two physical observables are considered equivalent if they coincide on shell.

Let us now illustrate the general formalism above by a few well-known examples from field theory.

\begin{ex}(Scalar field.) The dynamics of a single scalar field $\varphi$ in four-dimensional Minkowski  space are governed by the action
$$
S[\varphi]=\int  \Big(\frac12\partial_\mu \varphi\partial^\mu\varphi -V(\varphi)\Big)d^4x \,.
$$
Varying this action functional, we get the equation of motion and the presymplectic potential:
$$
\square \varphi +V'(\varphi)=0 \,, \qquad \Theta=\int_{\Sigma } \partial^\mu \varphi\delta \varphi \wedge  d^3 x_\mu\,.
$$
Note that only the terms with derivatives of $\varphi$ contribute to $\Theta$. The corresponding   presymplectic structure is
\begin{equation}\label{osc}
{{{\Omega}}}=\int_\Sigma \delta\varphi \wedge \partial^\mu\delta\varphi\wedge  d^3 x_\mu\,.
\end{equation}
If we chose $\Sigma$ to be a time slice $x^0=const$, then the presymplectic structure takes the form
\begin{equation}\label{PSsc}
{\Omega}=\int \delta\varphi \wedge \delta\dot \varphi \wedge d^3  \mathbf{x}\,.
\end{equation}
The corresponding equal-time Poisson brackets of fields and velocities read
$$
\{\varphi(\mathbf{x}),\varphi(\mathbf{x}')\}=0\,,\quad \{\dot\varphi(\mathbf{x}),\varphi(\mathbf{x}')\}=\delta^3(\mathbf{x}-\mathbf{x}')\,,\quad \{\dot\varphi(\mathbf{x}),\dot\varphi(\mathbf{x}')\}=0\,.
$$
This is in line with the usual Hamiltonian formalism  as for the scalar field $\dot \varphi=\pi$, where $\pi$ is the canonical momentum. Thus, the on-shell presymplectic structure (\ref{PSsc}) is nondegenerate and coincides with the canonical one.

\end{ex}

\begin{ex}(Spinor field.) Consider the Majorana spinor field $\psi$ of mass $m$ with the action
$$
S[\psi]=\int \left( i\bar\psi\bar{\sigma}^\mu\partial_\mu\psi -\frac m 2\psi\psi -\frac m2\bar\psi\bar\psi\right)d^4x\,.
$$
For the two-component spinor formalism and definition of $\sigma$-matrices see e.g. \cite{DS}.
Notice that the very possibility to write the mass term implies that the components of the Majorana spinor $\psi$ anticommute.

The equations of motion and the presymplecic structure following from this action are given by
\begin{equation}\label{osp}
i\bar{\sigma}^\mu\partial_\mu\psi=m \bar\psi\,,\qquad \Omega=\int_\Sigma i\delta\bar\psi\wedge \bar\sigma^\mu \delta\psi\wedge d^3x_\mu\,.
\end{equation}
For the Cauchy surface $\Sigma: x^0=const$, the equal-time Poisson brackets of component fields are
$$
\{\psi_\alpha(\mathbf{x}),\psi_\beta(\mathbf{x}')\}=0\,,\qquad \{\psi_\alpha(\mathbf{x}),\bar\psi_{\dot\alpha}(\mathbf{x}')\}=-i\sigma^0_{\alpha\dot\alpha}\delta^3(\mathbf{x}-\mathbf{x}')\,.
$$
Notice that the matrices of the presymplectic structure and Poisson brackets are purely imaginary for fermionic fields. Again, this is in harmony with the canonical formalism where the canonical momentum of the spinor field $\psi_\alpha$ is given by $\pi^\alpha=i(\bar\psi\bar{\sigma}^0)^\alpha $.

\end{ex}

 \begin{ex} (Gauge vector field.) The standard action functional for the free, massless, vector field $A=A_{\mu}dx^\mu$ is given by
$$
S[A]=
\frac14\int F_{\mu\nu}F^{\mu\nu}d^4x \,,\qquad F_{\mu\nu}=\partial_\mu A_\nu-\partial_\nu A_\mu\,.
$$
The action is clearly invariant under the gauge transformations
\begin{equation}\label{A-ginv}
\delta_\zeta A_\mu=\partial_\mu\zeta\,,
\end{equation}
where $\zeta$ is an arbitrary scalar function. The variation of this action yields the following field equations  and the presymplectic potential:
$$
\partial^\nu F_{\nu\mu}=0\,,\qquad  \Theta = \int_{\Sigma} F^{\mu\nu} \delta A_\mu \wedge d^3x_\nu\,.
$$
Because of the gauge invariance (\ref{A-ginv}) the corresponding  presymplectic form
\begin{equation}\label{OA}
\Omega = \int_{\Sigma} \delta F^{\mu\nu}\wedge \delta A_\mu\wedge  d^3x_\nu\,.
\end{equation}
is on-shell degenerate. Indeed,
$$
\begin{array}{rcl}
\Omega[\delta A,\delta_\zeta  A]&=&\displaystyle \int_\Sigma \delta F^{\mu\nu}\partial_\mu\zeta\wedge d^{3}x_\nu\\[5mm]
&=&\displaystyle \int_\Sigma \partial_\mu(\delta F^{\mu\nu}\zeta)\wedge d^3x_\nu-\int_\Sigma \zeta \delta(\partial_\mu F^{\mu\nu})\wedge d^3x_\nu\\[5mm]
&\approx&\displaystyle \int_{\partial \Sigma}\zeta \delta F^{\mu\nu}\wedge d^2x_{\mu\nu}=0\,.
\end{array}
$$
The last integral vanishes due to the zero boundary conditions for $A$.

 \end{ex}

\begin{ex} (General relativity.)
In the vierbein formulation of general relativity the space-time geometry is  described by a vierbein $e^a=e^a_\mu dx^\mu$ and a Lorentz connection $\omega^{ab}=\omega^{ab}_\mu dx^\mu$, $\omega^{ab}=-\omega^{ba}$.  The curvature tensor of the Lorentz connection has the standard form
$$
R_{\mu\nu}{}^a{}_b=\partial_\mu\omega_\nu{}^a{}_b -\partial_\nu\omega_\mu{}^a{}_b +[\omega_\mu,\omega_\nu]^a{}_b\,.
$$
Hereafter all the Lorentz indices are raised and lowered with the help of the Minkowski metric $\eta_{ab}$.

The Einstein-Hilbert action with cosmological term can be written as
\begin{equation}\label{HE}
S[e,\omega]=\frac12\int  \epsilon^{\mu\nu\lambda\delta}\epsilon_{abcd} e^a_\mu e^b_\nu\Big( R_{\lambda\delta}^{cd}+\frac12\Lambda e^c_\lambda e^d_\delta\Big)d^4x\,.
\end{equation}
Variation with respect to the connection $\omega$ yields
$$
D_\mu e^a_\nu-D_\nu e^a_\mu=0\,,
$$
where $D=d+\omega$ is the Lorentz-covariant derivative. This equation allows one to express  the Lorentz connection in terms of the vierbein, $\omega=\omega(e,\partial e)$.
Varying (\ref{HE}) with respect to $e$,  we get
$$
e^\nu_a R_{\mu\nu}^a{}_b=\Lambda e_{\mu b}\,,
$$
where $e^\nu_a$ is the inverse matrix of $e^a_\nu$. This equation is equivalent to the vacuum Einstein equation with cosmological term.

Using the general definition (\ref{Omega}), one can find the following presymplectic form on the space of vierbeins and Lorentz connections:
\begin{equation}\label{GR-PS}
\Omega= 2\int_\Sigma \epsilon^{\mu\nu\lambda\delta}\epsilon_{abcd} e^a_\mu\delta e^b_\nu\wedge \delta\omega^{cd}_\lambda\wedge d^3x_\delta\,.
\end{equation}
Notice that the cosmological term does not contribute to $\Omega$.

Besides the space-time diffeomorphisms, action (\ref{HE}) is invariant under the local Lorentz transformations
$$
\delta_\zeta e^a=\zeta^{ab}e_b\,,\qquad \delta_\zeta \omega^{ab} =D\zeta^{ab}\,,
$$
with $\zeta^{ab}=-\zeta^{ba}$ being infinitesimal gauge parameters.  As a result, the presymplectic form (\ref{GR-PS}) is necessarily degenerate on shell.

Unlike the previous example, the Einstein-Hilbert action is rather nonlinear and so is the presymplectic structure (the components of $\Omega$ depend on the vierbein field $e$).   Linearization about  a particular vacuum solution $e=h$, $\omega=w$, e.g. (anti-)de Sitter one, brings the presymplectic structure (\ref{GR-PS}) into the form
\begin{equation}\label{s2}
\hat{\Omega}= 2\int_\Sigma \epsilon^{\mu\nu\lambda\delta}\epsilon_{abcd} h^a_\mu\delta \hat{e}^b_\nu\wedge \delta\hat{\omega}^{cd}_\lambda\wedge d^3x_\delta\,,
\end{equation}
where $\hat{e}$ and $\hat{\omega}$ are fluctuations around the background vierbein $h$ and  the Lorentz connection $w$. This presimplectic form is also degenerate along the directions of the linearized gauge transformations.
\end{ex}

 More examples of presymplectic structures, including those associated with the frame-like Lagrangians for massless higher-spin fields, can be found in \cite{AG}.

In the following it will be convenient to work with the nonintegrated density $\mathit{\Omega}$ of the presymplectic form $\Omega$. This is defined as
\begin{equation}\label{PS}
\Omega = \int_\Sigma\mathit{\Omega}\,.
\end{equation}
Here the integrand $\mathit\Omega =\mathit\Omega(\delta\phi,\delta\phi)$ may be viewed as an antisymmetric, bidifferential operator in $\delta\phi$'s with coefficients depending on the derivatives of $\phi$'s and taking values in ($n-1$)-forms on the space-time manifold. Another way of thinking  of $\mathit{\Omega}$ is to interpret it as a ``hybrid'' differential form of type $(2,n-1)$, that is, a variational $2$-form in the functional space $\Phi$ and a usual ($n-1$)-form on the space-time manifold. This last point of view can be  formalized within the concept of variational bicomplex, see e.g. \cite{Anderson}, \cite{Sh1}, although we  will not dwell on it here. In the sequel we will refer to $\mathit\Omega$ as a \textit{presymplectic current}.  In order to have a one-to-one correspondence between the covariant presymplectic structures (\ref{PS}) and their currents we will identify two presymplectic currents $\mathit\Omega$ and $\mathit\Omega'$  if they differ by a $d$-exact $(2, n-1)$-form, i.e.,
 $$
 \mathit{\Omega}'-\mathit{\Omega}=d\beta
 $$
for some $(2,n-2)$-form $\beta$. In this case we write $\varOmega'\simeq \varOmega$. Besides, we assume the space $\Phi$ to consist of fields that vanish at space infinity.

The notions of presymplectic form and presimplectic current can be extended beyond the scope of variational dynamics. Let we have given a set of (not necessarily Lagrangian) field equations
\begin{equation}\label{E}
E_a(\phi,\partial\phi, \ldots,\partial^n\phi)=0
\end{equation}
for a collection of fields $\phi^i$ living on an $n$-dimensional manifold $N$. Following \cite{BHL}, \cite{Kh}, we say that a $(2,n-1)$-form $\mathit \Omega$ on $\Phi\times N$ defines a presymplectic current compatible with equations of motion (\ref{E}) if
\begin{equation}\label{PF}
\delta \mathit{\Omega}\simeq 0\,,\qquad d\mathit{\Omega} \approx 0\,.
\end{equation}
The form $\mathit \Omega$ is assumed to have the following appearance:
\begin{equation}\label{vOm}
\mathit{\Omega}=\sum_{k,l=0}^{K,L}\mathit{\Omega}_{ij\mu_1\cdots\mu_{n-1}}^{\nu(k),\lambda(l)}
\delta\phi^i_{\nu(k)}\wedge \delta\phi^j_{\lambda(l)}\wedge dx^{\mu_1}\wedge \cdots\wedge dx^{\mu_{n-1}}\,,
\end{equation}
where
$$
\phi^i_{\nu(k)}=\phi^i_{\nu_1\nu_2\cdots\nu_k}=\partial_{\nu_1}\partial_{\nu_2}\cdots\partial_{\nu_k}\phi^i
$$
and the coefficients $\mathit{\Omega}_{ij\mu_1\cdots\mu_{n-1}}^{\nu(k),\lambda(l)}$ are given  by  smooth functions of fields and their derivatives up to some finite order.

The second equation in (\ref{PF}) ensures the independence of the corresponding presymplectic structure of the choice of the Cauchy surface $\Sigma$. Therefore, it induces a $2$-form on the phase space of all solutions to the field equations (\ref{E}). Then the first condition in (\ref{PF}) identifies this $2$-form as a presymplectic one. This presymplectic structure gives rise to a Poisson bracket in the space of Hamiltonian $(0,n-1)$-forms on $\Phi\times N$ in a similar way to the finite-dimensional presymplectic manifolds of Sec. \ref{PM}.  For a more detailed  discussion of the covariant phase space and the Poisson algebra of physical observables we refer the reader to \cite{Sh1}.

\section{Free HS fields in $AdS_4$}\label{HSFF}

The four-dimensional anti-de Sitter space, $AdS_4$, is a maximally symmetric solution to the vacuum Einstein equations with negative cosmological constant. As was first noticed by Fradkin and Vasiliev \cite{FV1}, \cite{FV2} it is the only background geometry with maximal symmetry that admits a consistent interaction of higher-spin massless fields. In the vierbein approach the geometry of $AdS_4$ is described by the vierbein $h^a=h^a_\mu dx^\mu$ and the compatible Lorentz connection $w^{ab}=w^{ab}_\mu dx^\mu$.   All Lorentz indices are raised and lowered by the Minkowski metric $\eta_{ab}$.

In the unfolded formulation of the free HS dynamics the massless particles of all spins in $AdS_4$ are described in terms of two master fields: the \textit{gauge field} $\omega$ and the \textit{Weyl field} $C$. These are given, respectively, by $1$- and $0$-form on $AdS_4$ with values in an infinite-dimensional associative algebra $W$,  the \textit{Weyl algebra}. It is the infinite-dimensionality of $W$ which allows the master fields to accommodate the whole spectrum of spins, from zero to infinity. Since the Weyl algebra is in the core of the HS field equations, we begin with explaining of its structure.  A comprehensive account of the subject can be found in  \cite{V0}, \cite{DS}.

As a linear space the algebra $W$ is given by formal power series in the complex variables $y^\alpha$, $\bar y^{\dot \alpha}$, $\alpha,\dot\alpha=1,2$, so that the general element of $W$ reads
 \begin{equation}\label{exp}
 f=\sum_{m, n}\frac{1}{m!n!}f_{\alpha(m)\dot\alpha(n)}(y^\alpha)^m (\bar y^{\dot\alpha})^n,\qquad f_{\alpha(m)\dot\alpha(n)}\in \mathbb{C}\,.
 \end{equation}
  As above we use the shorthand notation for symmetric indices, $$f_{\alpha(m)\dot\alpha(n)}=f_{\alpha_1\cdots\alpha_m\dot\alpha_1\cdots\dot\alpha_n}\,\qquad (y^\alpha)^m=y^{\alpha_1}\cdots y^{\alpha_m}\,.$$
  The mutually conjugate complex variables $y^\alpha$ and $\bar y^{\dot \alpha}$ may viewed as components of the left-handed and right-handed Weyl spinors for the Lorentz algebra $so(3,1)\sim sl(2,\mathbb{C})$.

  The multiplication in $W$ is given by the so-called Weyl-Moyal $\ast$-product:
$$
  f\ast g=\exp\left(i\epsilon^{\alpha\beta}\frac{\partial}{\partial y^\alpha}\frac{\partial}{\partial z^{\beta}} +i\epsilon^{\dot\alpha\dot\beta}\frac{\partial}{\partial \bar y^{\dot\alpha}}\frac{\partial}{\partial
    \bar z^{\dot\beta}}\right)f(y,\bar y)g(z,\bar z )|_{z=y}\,.
  $$
  Here $\epsilon^{\alpha\beta}$ and $\epsilon^{\dot\alpha\dot\beta}$ are the pair of $sl(2,\mathbb{C})$-invariant tensors defined by the rule
  $$
  \epsilon^{\alpha\beta}=-\epsilon^{\beta\alpha}\,, \qquad \epsilon^{12}=1\,,
  $$
 and the same for $\epsilon^{\dot\alpha\dot\beta}$. All spinor indices are raised and lowered by $\epsilon^{\alpha\beta}$, $\epsilon^{\dot\alpha\dot\beta}$ and their inverse:  $y^\alpha=\epsilon^{\alpha\beta}y_\beta$, $y_\alpha=y^\beta\epsilon_{\beta\alpha}$  and the same for dotted indices.   The two main properties of the $\ast$-product above are associativity and unitality, meaning that
 $$
 (f\ast g)\ast h=f\ast (g\ast h)\,,\qquad 1\ast f=f=f\ast 1\,,\qquad \forall f,g,h\in {W}\,.
 $$

 One more important property of the Weyl algebra, explaining to some extent its relevance to the HS dynamics, is that the Lie algebra of internal derivations of ${W}$ contains the anti-de Sitter algebra $so(3,2)\sim \mathrm{sp}(4)$ as a finite-dimensional subalgebra. This is generated by all the quadratic monomials
$$
M_{\alpha\beta}=-\frac{i}{2}y_\alpha y_\beta\,,\qquad \bar M_{\dot\alpha\dot\beta}=-\frac{i}{2}\bar y_{\dot\alpha} \bar y_{\dot\beta}\,,\qquad P_{\alpha\dot\alpha}=-\frac{i}{2}y_\alpha \bar y_{\dot\alpha}
$$
with respect to the $\ast$-commutator.
The monomials $M_{\alpha\beta}$ and $\bar M_{\dot\alpha\dot\beta}$ span the Lorentz subalgebra $so(3,1)\subset so(3,2)$ and $P_{\alpha\dot\alpha}$ correspond to the anti-de Sitter translations. Using the standard vector-spinor dictionary (see e.g. \cite{DS}) we can also pass to a more familiar basis of $so(3,2)$ generators labeled by the Lorentz indices:
\begin{equation}\label{PM}
P_a=\sigma_a^{\alpha\dot\alpha}P_{\alpha\dot\alpha}       \,, \qquad   M_{ab}=\sigma_{ab}^{\alpha\beta} M_{\alpha\beta}+\bar{\sigma}_{ab}^{\dot\alpha\dot\beta} M_{\dot\alpha\dot\beta}\,.
\end{equation}
In this basis the commutation relations take the form
$$
\begin{array}{l}
[M_{ab},M_{cd}]_\ast=\eta_{bc} M_{ad}-\eta_{ac}M_{bd}-\eta_{bd}M_{ac}+\eta_{ad}M_{bc}\,,\\[3mm]
[M_{ab},P_c]_\ast=\eta_{bc} P_a- \eta_{ac} P_b\,, \\[3mm]
[P_a,P_b]_\ast= M_{ab}\,.
\end{array}
$$

Tensoring now the Weyl algebra ${W}$ with the exterior algebra $\Lambda=\bigoplus_k\Lambda^k$ of differential forms on $AdS_4$, we arrive at the associative algebra $\mathcal F=\Lambda\otimes {W}$ with the $\star$-product defined by the rule
$$
(\alpha\otimes f)\star(\beta\otimes g)=(\alpha\wedge \beta)\otimes (f\ast g)\qquad \forall \alpha,\beta\in \Lambda\,,\quad \forall f,g\in {W}\,.
$$
The $\star$-product algebra is naturally graded with respect to the form degree: $$\mathcal{F}=\bigoplus_{p=0}^4\mathcal{F}^p\,,\qquad \mathcal{F}^p=\Lambda^p\otimes W\,, \qquad \mathcal{F}^p\star \mathcal{F}^q\subset \mathcal{F}^{p+q} \,.$$
We will denote the form-degree of a homogeneous element $F\in \mathcal{F}^p$ by $|F|=p$.

Geometrically, the elements of $\mathcal{F}$ may be viewed as differential forms with values in $W$:
$$
F(y,\bar y|x,dx)=\sum_{m,n} \frac1{m!n!}F_{\alpha(m)\dot\alpha(n)}(x,dx)(y^{\alpha})^m(\bar y^{\dot\alpha})^n\,.
$$
The expansion coefficients $F_{\alpha(m)\dot\alpha(n)}(x,dx)$ are then naturally interpreted  as form-valued, spin-tensor fields on $AdS_4$. In accordance with the standard  relationship between spin and statistics the component fields with even number of spinor indices are treated as bosons, while those with odd number of spinor indices are declared to be fermions.

Associated to $\mathcal{F}$ is the Lie superalgebra $\mathcal{L}(\mathcal{F})$ with the $\star$-commutator
$$
[F,G]_\star=F\star G-(-1)^{|F||G|}G\star F\,.
$$
Considering the first factor in the tensor product $\mathcal{F}=\Lambda\otimes {W}$ as basic ring, one can endow the algebra $\mathcal{F}$ with a supertrace operation $\mathrm{Str}: \mathcal{F}\rightarrow \Lambda$ and a nondegenerate  inner product $\langle \,\cdot\,|\,\cdot\, \rangle: \mathcal{F}\otimes \mathcal{F}\rightarrow \Lambda$. These are given by \cite{V0}
\begin{equation}\label{InPr}
\mathrm{Str}(F)=F(0,0|x,dx)\,,\qquad \langle F|G \rangle=\mathrm{Str}(F\star G)\,.
\end{equation}
By definition,  the supertrace vanishes on the $\star$-commutators,
$$
\mathrm{Str}([F,G]_\star)=0 \qquad \forall F,G\in \mathcal{F}\,,
$$
and the inner product is invariant under the adjoint action of $\mathcal{L}(\mathcal{F})$:
$$
\langle [H,F]_\star|G\rangle+(-1)^{|H||F|}\langle F|[H,G]_\star\rangle=0\,\qquad \forall H,F,G\in \mathcal{F}\,.
$$
For later use we also define the $\star$-anticommutator
$$
\{F,G\}_\star=F\star G +(-1)^{|F||G|} G\star F\,.
$$

The background geometry of $AdS_4$ allows us to endow the algebra $\mathcal{F}$ with the Lorentz covariant differential $D: \mathcal{F}^p\rightarrow \mathcal{F}^{p+1}$. This is defined in terms of the Lorentz connection $w^{ab}$ as
$$
DF=dF+[w, F]_\star\,,\qquad w=\frac12w^{ab} M_{ab}\,,
$$
with $M_{ab}$ being given by (\ref{PM}).

By definition, the operator  $D$ differentiates the $\star$-product by the Leibniz rule,
$$
D(F\star G)=(DF)\star G+(-1)^{|F|}F\star DG\qquad \forall F,G\in \mathcal{F}\,,
$$
but is not nilpotent; instead,  we have the curvature $2$-form
\begin{equation}\label{R}
D^2F=[R, F]\,,\qquad R=\frac\Lambda 3 h^a\wedge h^b M_{ab} \in \mathcal{F}_2\,,
\end{equation}
which is proportional to the cosmological constant $\Lambda$.  In the following we set $\Lambda=-3$.

Besides the covariant differential and curvature, the algebra $\mathcal{F}$ possesses a distinguished element associated to the $AdS_4$ vierbein, namely, $h=h^aP_a\in \mathcal{F}_1$. The compatibility of the Lorentz connection and vierbein implies that  the $1$-form $h$ is covariantly constant, that is,
\begin{equation}\label{Dh}
Dh=0\,.
\end{equation}
 By making use of $h$, we can write the curvature $2$-form as the $\star$-square of the vierbein form
 \begin{equation}\label{Rhh}
 R=-h\star h\,.
 \end{equation}
 Taken together the last two equations imply the Bianchi identity $D R=0$ for the curvature. It is worthy of note that all three Eqs. (\ref{R}), (\ref{Dh}), and (\ref{Rhh}) are equivalent to a single zero-curvature condition for the anti-de Sitter covariant differential
\begin{equation}\label{DD}
\mathcal{D}=D+[h,\,\cdot\,]_\star\,,\quad \mathcal{D}^2=0\,.
\end{equation}

Now we are ready to present the free HS equations for the master fields $C\in \mathcal{F}^0$ and $\omega\in \mathcal{F}^1$. These read \cite{V0}, \cite{DS}
\begin{equation}\label{HSE}
D\omega +[h, \omega]_\star=\hat{H}_+C_-+\hat{H}_-C_+\,,\qquad
DC+\{h,C\}_\star=0\,.
\end{equation}
Here we introduced the projections onto the purely holomorphic and anti-holomorphic sectors of the Weyl field,
$$
C_+=C(y,0|x)\,,\qquad C_-=C(0,\bar y|x)\,,
$$
together with the pair of differential operators
$$
\hat{H}_+=H^{\dot\alpha\dot\beta}\partial_{\dot\alpha}\partial_{\dot\beta}\,,\qquad \hat{H}_-=H^{\alpha\beta}\partial_\alpha\partial_\beta\,,
$$
$$
H^{\dot\alpha\dot\beta}=h^{\gamma\dot \alpha}\wedge h_{\gamma}{}^{\dot \beta}\,,\quad H^{\alpha\beta}=h^{\alpha\dot\gamma}\wedge h^{\beta}{}_{\dot \gamma}\,.
$$
Notice that the left hand side of the first equation in (\ref{HSE}) is given by the anti-de Sitter  covariant differential (\ref{DD}) of the gauge field $\omega$.

Since the system (\ref{HSE}) is linear, it decouples into an infinite set of subsystems for the
particles of definite spin $s =0, 1/2, 1,\ldots$ . In order to single out the contribution of each particular spin, it is convenient to introduce the pair of operators
$$
\hat{N}_\pm=y^\alpha\partial_\alpha\pm\bar y^{\dot \alpha}\partial_{\dot\alpha}\,,
$$
which count the number of $y$'s and $\bar y$'s. Then the master fields decompose into the sums
$$
\omega=\sum _{s=1}^\infty\omega_{^{(s)}}\,,\qquad C=\sum_{s=0}^\infty (C_{^{(s)}}+\bar C_{^{(s)}})\,,
$$
where
$$
\hat{N}_+\omega_{^{(s)}} =2(s-1)\omega_{^{(s)}}\,,\qquad \hat{N}_-C_{^{(s)}}=2s C_{^{(s)}}\,,\qquad \hat{N}_-\bar C_{^{(s)}}=-2s \bar C_{^{(s)}}\,.
$$
The massless particle of spin  $s$ is now described by the component
fields $(\omega_{^{(s)}}, C_{^{(s)}})$ with the understanding that $\omega_{^{(0)}}=\omega_{^{(1/2)}}=0$ (no gauge fields for the scalar and spin-$1/2$ particles). Being expanded in $y$'s and $\bar y$'s,
these fields generate an infinite number of spin-tensor fields on
$AdS_4$:
\begin{equation}\label{s-exp}
\omega_{^{(s)}}=\{\omega_{\alpha(2s-2-n)\dot\alpha(n)}\}_{n=0}^{2s-2}\,,\qquad
C_{^{(s)}}=\{C_{\alpha(2s+n)\dot\alpha(n)}\}_{n=0}^\infty\,.
\end{equation}
Most of these fields are auxiliary and can be in principle excluded  from consideration by means of the equations of motion.

The system of field equations (\ref{HSE}) is known to be non-Lagrangian.  This fact can be seen as follows.
 First, we note that the equations (\ref{HSE}) are not independent, rather they obey a set of gauge identities that follow from applying the covariant differential $D$ to both the sides of (\ref{HSE}).  This property is usually referred to as the formal consistency of the HS system  (no hidden integrability conditions).  The space of gauge identities is naturally parameterized by the elements of $\mathcal{F}^1\oplus \mathcal{F}^2$. Besides the gauge identities, the system (\ref{HSE}) enjoys the obvious gauge invariance
\begin{equation}\label{GSym}
\delta_\zeta \omega =\mathcal{D}\zeta\,,\qquad\delta_\zeta C=0\,,\qquad \forall \zeta\in \mathcal{F}^0\,.
\end{equation}
If the system (\ref{HSE}) were Lagrangian then, according to the second Noether's theorem \cite{KS}, there would be a one-to-one correspondence between the gauge symmetries and identities, which is not the case as $\mathcal{F}^0 \neq\mathcal{F}^1\oplus \mathcal{F}^2$.   Furthermore, the gauge identities appear to be reducible, while the gauge symmetries (\ref{GSym}) are not. The last disagreement is also impossible for Lagrangian systems.

It is the non-Lagrangian nature of the free HS equations (\ref{HSE}) and their nonlinear extensions, see \cite{V0}, \cite{DS},  which presents a real challenge to the quantization of HS theories.

Let us now explain how the system  (\ref{HSE}) works in the case of low spins.

\vspace{3mm}
\noindent
${{{\textbf{Spin 0}}}.}$ As is seen from (\ref{s-exp}) no gauge fields correspond to the scalar massless  particle; all the fields are accommodated in $C_{^{(0)}}=\{C_{\alpha(n)\dot\alpha(n)}\}_{n=0}^\infty$. The first term of this sequence is identified with the scalar field itself
$$
\varphi(x)=C(0,0|x)\,.
$$
This takes values in the center of the Weyl algebra. The second equation in (\ref{HSE}) gives rise to an infinite chain of equations for the fields $C_{^{(0)}}$.  The chain starts with the equation
\begin{equation}\label{1es}
d\varphi=-ih^{\alpha\dot\alpha}C_{\alpha\dot \alpha } \quad \Leftrightarrow \quad C_{\alpha\dot\alpha}=\frac i2 D_{\alpha\dot\alpha} \varphi\,,
\end{equation}
where $D_{\alpha\dot\alpha}=\sigma^a_{\alpha\dot\alpha}h_a^\mu D_\mu$ and $h_a^\mu$ is the inverse vierbein. In fact, $C_{\alpha\dot\alpha}$ is just the notation for the first partial derivatives of $\varphi$. The next  equation is given by
$$
DC_{\alpha\dot\alpha}=-ih^{\beta\dot\beta}C_{\alpha\beta\dot\alpha\dot\beta}+ih_{\alpha\dot\alpha}\varphi\,.
$$
Converting the world indices into the Lorentz ones by the inverse vierbein and using (\ref{1es}), one can find  that this equation is equivalent to
$$
(\Box - 8)\varphi=0\,,\qquad C_{\alpha\beta\dot\alpha\dot\beta}=\frac i2D_{(\alpha\dot\alpha} C_{\beta\dot\beta{})}=-\frac14 D_{(\alpha\dot\alpha}D_{\beta\dot\beta)}\varphi\,,
$$
where $\Box$ is the d'Alembert operator on $AdS_4$ and the round
brackets mean symmetrization of doted and undoted indices. We see
that the scalar field $\varphi$ does satisfy the Klein-Gordon equation for the
massless particle on $AdS_4$, while  the spin-tensors
$C_{\alpha\dot\alpha}$ and $C_{\alpha\beta\dot\alpha\dot\beta}$ play the
role of auxiliary fields. The same mechanism works for all higher
spin-tensor fields: the second equation in (\ref{HSE}) can be solved for
$C_{\alpha(n)\dot\alpha(n)}$ in terms of the successive derivatives
of $\varphi$, bringing no new constraints on the dynamical field
$\varphi$ itself. This way of formulating field dynamics trough an infinite system of the first-order equations is known
as the \textit{unfolded representation} \cite{V0}, \cite{DS}.

\vspace{3mm}
\noindent
${{\mathbf{Spin \;1/2}}.}$ The massless spin-$1/2$ particle is described by the sequence of spin-tensor fields $C_{^{(1/2)}}=\{C_{\alpha(n+1)\dot\alpha(n)}\}_{n=0}^\infty$. As there is no gauge symmetry, the first equation in (\ref{HSE}) is absent, while the second one gives an infinite chain of equations starting with
$$
DC_\beta =-ih^{\alpha\dot\alpha}C_{\beta\alpha\dot\alpha}\,.
$$
The Dirac equation is a simple consequence of this one,
$$
D_{\alpha\dot\alpha}C^\alpha=0\,,
$$
cf. (\ref{osp}). All the rest equations impose no restrictions on the dynamical field $\psi^\alpha=C^\alpha$, just expressing the higher spin-tensors via the covariant derivatives of $\psi^\alpha$.

\vspace{3mm}
\noindent
${{\mathbf{Spin \;1}}.}$ The gauge potential of the electromagnetic field is identified with
$$
A=\omega(0,0|x,dx)\,.
$$
The gauge transformations (\ref{GSym}) imply that
$$
\delta_\zeta A=d\zeta
$$
for arbitrary scalar field $\zeta$. The first equation in (\ref{HSE}) just provides the standard  connection between the gauge potential and the strength tensor of the electromagnetic field
\begin{equation}\label{dA}
dA=h^{\alpha\dot\alpha}\wedge h^{\beta}{}_{\dot\alpha}C_{\alpha\beta}+h^{\alpha\dot\alpha}\wedge h_{\alpha}{}^{\dot\beta}\bar C_{\dot\alpha\dot\beta}\,.
\end{equation}
From the second equation in (\ref{HSE}) we then deduce that
$$
D_{\beta\dot\alpha}C^{\beta}{}_{\alpha}=0\,,\qquad D_{\alpha\dot\beta}\bar C^{\dot\beta}{}_{\dot\alpha}=0\,.
$$
This is nothing but the spinorial  version of the standard Maxwell's equations for the free electromagnetic field on the anti-de Sitter background.

All the other spin-tensor fields $C_{\alpha(n+2)\dot\alpha(n)}$  and $\bar C_{\alpha(n)\dot\alpha(n+2)}$ with $n>0$ appear to be auxiliary and can be expressed through the successive derivatives of the strength tensor (\ref{dA}) in perfect analogy to the cases of scalar and spin-$1/2$ fields.

\vspace{3mm}
\noindent
${{\textbf{Spin 2}}.}$ The sector of gauge fields $\omega_{^{(2)}}$ includes now the $1$-forms
$$
\omega_{\alpha\beta},\qquad \omega_{\alpha\dot\beta}\,,\qquad \bar\omega_{\dot\alpha\dot\beta}\,.
$$
These are naturally identified with the fluctuations of the background vierbein $h^{\alpha\dot\alpha}$ and the Lorentz connection $w^{\alpha\beta}$, $\bar w^{\dot\alpha\dot\beta}$.
Compatibility between the full Lorentz connection and the  vierbein implies a certain relation between the aforementioned fluctuations. This relation is exactly reproduced by the first equation in (\ref{HSE}), namely,
$$
D\omega^{\alpha\dot\alpha}+\omega^{\alpha}{}_\gamma \wedge h^{\gamma\dot\alpha}- h^{\alpha\dot\gamma}\wedge \bar\omega_{\dot\gamma}{}^{\dot\alpha} =0\,.
$$
Besides, it gives the following relations between the components of the gauge and Weyl fields:
\begin{equation}\label{CC}
\begin{array}{l}
D\omega^{\alpha\beta} +h^{\alpha}{}_{\dot\gamma}\wedge \omega^{\beta\dot\gamma}+ h^{\beta}{}_{\dot\gamma}\wedge \omega^{\alpha\dot\gamma} =h_{\gamma\dot\delta}\wedge h_{\delta}{}^{\dot\delta} C^{\alpha\beta\gamma\delta}\,,\\[3mm]
D\bar\omega^{\dot\alpha\dot\beta} +h_{\gamma}{}^{\dot\alpha}\wedge \omega^{\gamma\dot\beta}
+h_{\gamma}{}^{\dot\beta}\wedge \omega^{\gamma\dot\alpha} =h_{\delta\dot\gamma}\wedge h^{\delta}{}_{\dot\delta} \bar C^{\dot\alpha\dot\beta\dot\gamma\dot\delta}\,.
\end{array}
\end{equation}
The left hand sides of these equations involve the linearized curvature  of the Lorenz connection plus  terms proportional to the cosmological constant.

The role of these equations is twofold. First,  they identify the spin-tensor fields $C^{\alpha\beta\gamma\delta}$ and  $\bar C^{\dot\alpha\dot\beta\dot\gamma\dot\delta}$ as the components of the linearized Weyl tensor associated to the curvature tensor. (This justifies the name ``Weyl field'' for $C$.) Second, they imply that the linearized Einstein's equations with cosmological constant hold for $\omega$'s.

It should be noted that equations (\ref{CC}) are consistent provided that the Weyl tensor obeys the Bianchi identities
 $$
 D_{\alpha\dot\alpha} C^{\alpha}{}_{\beta\gamma\delta}=0\,,\qquad D_{\alpha\dot\alpha}\bar C^{\dot\alpha}{}_{\dot\beta\dot\gamma\dot\delta}=0\,.
 $$
 These identities are encoded by the second equation in (\ref{HSE}). Again, one can see that similar to the Weyl tensor, all the higher spin-tensor fields  $C_{\alpha(n+4)\dot\alpha(n)}$ and $\bar C_{\alpha(n)\dot\alpha(n+4)}$ play an auxiliary role and can be consistently excluded from the theory by means of the second equation.

\section{Presymplectic currents for free HS fields}\label{CPS}

In this section, we propose a family of covariant presymplectic
structures for the non-Lagrangian field equations (\ref{HSE}). Since the HS
fields we are dealing with are free, it is natural to look for
the presymplectic structure (\ref{vOm}) whose components are independent of
fields.  Then  the presymplectic current  in
question can be written schematically as
\begin{equation}\label{KLN}
\mathit{\Omega} =K\delta \omega\wedge \delta \omega +L\delta
\omega\wedge \delta C + N\delta C\wedge \delta C\,,
\end{equation}
where $K$, $L$, and $N$ are some field-independent $3$-forms on
$AdS_4$ with values in bidifferential operators acting on the
variational differentials of the master fields. Being field
independent, the form (\ref{KLN}) is automatically $\delta$-closed
and the only nontrivial condition to satisfy is the on-shell
clossedness with respect to the de Rham differential $d$. The
anti-de Sitter invariance of the form (\ref{KLN}) -- a necessary
condition for the relativistic symmetry to survive quantization --
imposes further restrictions on the structure coefficients $K$, $L$,
and $N$.

In order to facilitate the analysis of expression (\ref{KLN}) we will
make use of the natural grading in the field space. According to the
expansion (\ref{exp}) the Weyl algebra decomposes into
the direct sum  $W=\bigoplus W_{n,m}$ of finite-dimensional
subspaces spanned by homogeneous polynomials in $y$'s and $\bar
y$'s,
$$
\hat{N}_\pm f=(n\pm m)f\quad \Leftrightarrow\quad f \in W_{n,m}\,.
$$
This bigraduation extends immediately to the space
$\mathcal{F}=\bigoplus \mathcal{F}_{n,m}$ by setting
$\mathcal{F}_{n,m}=\Lambda\otimes W_{n,m}$ and one can easily see
that the different homogeneous subspaces $\mathcal{F}_{n,m}$ are orthogonal to
each other with respect to the inner product (\ref{InPr}),
$$
\langle F| G\rangle \propto \delta_{m,k}\delta_{n,l}\qquad \forall F\in \mathcal{F}_{m,n}\,,\quad \forall G\in \mathcal{F}_{k,l}\,.
$$

Expanding now the master fields into homogeneous components,
$$
\omega=\sum_{m,n}\omega_{mn}\,,\qquad C=\sum_{m, n}C_{mn}\,,
$$
we can bring the HS equations (\ref{HSE}) into the form
$$
D\omega_{mn}+\hat{h}_+\omega_{m-1,n+1}+\hat{h}_-\omega_{m+1,n-1}=\delta_{m,0}\hat{H}_+C_{0,n+2}+\delta_{0,n}\hat{H}_-C_{m+2,0}\,,
$$
\begin{equation}\label{EoM}
DC_{mn}+\hat{h}_0C_{m+1,n+1}+2h\cdot C_{m-1,n-1}=0\,,\qquad m,n=0,1,
\ldots\,,
\end{equation}
where the operators $\hat{h}_\pm$ and $\hat{h}_0$ are defined by the relations
$$
\hat{h}=[h,\cdot\,]_\star =\hat{h}_++\hat{h}_-\,,\qquad
\hat{h}_+=h^{\alpha\dot\alpha}y_\alpha\partial_{\dot\alpha}\,,\quad
\hat{h}_-=h^{\alpha\dot\alpha}y_{\dot\alpha }\partial_\alpha\,,
$$
and
$$
\hat{h}_0=ih^{\alpha\dot\alpha}\partial_\alpha\partial_{\dot\alpha}\,,\qquad h=-\frac i2h^{\alpha\dot\alpha}y_\alpha\bar y_{\dot\alpha}\,.
$$
The introduced operators possess the
following properties:
$$\hat{h}^2_{\pm}=\hat{h}^2_0=0\,, \qquad [D,\hat{h}_\pm]=[D,\hat{h}_0]=Dh=0\,,\qquad \hat{h}_\pm \hat{H}_\pm=\hat{h}_0 \hat{H}_\pm=0\,. $$
\begin{equation}\label{id}
\langle A|\hat{h}_+ B\rangle=-(-1)^{|A|}\langle \hat{h}_- A|B\rangle\qquad\forall A,B\in \mathcal{F} \,.
\end{equation}
The last relation is enough to check for homogeneous $A$ and $B$.

Let us now introduce the following set   $\hat{{\varOmega}}=\{\mathit{\Omega}_{mn}\}_{m,n=0}^\infty $ of complex $(2,3)$-forms:
\begin{equation}\label{Omn}
\mathit{\Omega}_{mn}=\left\{
                       \begin{array}{ll}
                         \langle\delta\omega_{mn}|\hat{h}_+\delta\omega_{m-1,n+1}\rangle, & \hbox{for $m>0$;} \\[3mm]
                         -\langle\delta \omega_{0,n}|\hat{H}_+\delta C_{0,n+2}\rangle, & \hbox{for $m=0$.}
                       \end{array}
                     \right.
\end{equation}
Considering  ${\hat{\varOmega}}$ as an infinite square matrix, we first  observe that the matrix  $\hat{\varOmega}$ is anti-Hermitian modulo equations of motion and $d$-exact forms.
Namely,
\begin{equation}\label{Herm}
{\varOmega}_{mn}+\bar{\varOmega}_{nm} \approx d\Psi_{mn}\,,\qquad\Psi_{mn}=-\frac12\langle\delta\omega_{mn}|\delta\omega_{mn}\rangle\,.
\end{equation}
Then using Rel. (\ref{id})\footnote{Note that $|A|$ is given now by the \textit{total form degree}  which  counts the differentials of the space-time coordinates \textit{and} the variational differentials of bosonic fields. The variational differentials of fermionic fields are Grassmann even, and hence commute.}, we find that for $m>0$
$$
\begin{array}{rcl}
\bar\varOmega_{mn}&=&\langle\delta\omega_{nm}|\hat{h}_-\delta\omega_{n+1,m-1}\rangle
=-\langle\hat{h}_+\delta\omega_{nm}|\delta\omega_{n+1,m-1}\rangle\\[3mm]
&=&-\langle\delta\omega_{n+1,m-1}|\hat{h}_+\delta\omega_{nm}\rangle=-\varOmega_{n+1,m-1}\,.
\end{array}
$$
Taken together with (\ref{Herm}), the last equality implies that
\begin{equation}\label{ER}
\mathit{\Omega}_{mn}\simeq \mathit{\Omega}_{m-1,n+1}\qquad \forall m>0\,,
\end{equation}
where the sign $\simeq$ means equality modulo equations of motion {and} $d$-exact forms.

Now we claim that the entries of the matrix $\hat{\varOmega}$ are given by on-shell closed forms. In view of the equivalence relation (\ref{ER}) it is enough to check the statement only for  representatives of the equivalence classes, e.g. the forms $\mathit{\Omega}_{m,0}$. Let us first assume that $m>0$, then
$$
d\mathit{\Omega}_{m,0}=d\langle \delta\omega_{m,0}| \hat{h}_+\delta\omega_{m-1,1}\rangle
$$
$$
=-\langle \delta D\omega_{m,0}| \hat{h}_+\delta\omega_{m-1,1}\rangle+\langle \delta\omega_{m,0}| \hat{h}_+\delta D\omega_{m-1,1}\rangle
$$
$$
=-\langle \delta D\omega_{m,0}| \hat{h}_+\delta\omega_{m-1,1}\rangle-\langle \hat{h}_-\delta\omega_{m,0}|\delta D\omega_{m-1,1}\rangle
$$
$$
\approx-\langle \hat{h}_+\delta \omega_{m-1,1}+\hat{H}_-\delta C_{m+2,0}|\hat{h}_+\delta\omega_{m-1,1}\rangle
$$
$$
-
\langle \hat{h}_-\delta\omega_{m,0}|\hat{h}_-\delta\omega_{m,0}+\hat{h}_+\delta\omega_{m-2,2}\rangle
$$
$$
=-\langle\hat{H}_-\delta C_{m+2,0}|\hat{h}_+\delta\omega_{m-1,1}\rangle
-\langle \hat{h}_-\delta\omega_{m,0}|\hat{h}_+\delta\omega_{m-2,2}\rangle
$$
$$
=-\langle \hat{h}_-\hat{H}_-\delta C_{m+2,0}|\delta\omega_{m-1,1}\rangle
+\langle \delta\omega_{m,0}|\hat{h}^2_+\delta\omega_{m-2,2}\rangle=0\,.
$$
Here we used the equations of motion (\ref{EoM}) and identities (\ref{id}). In case $m=0$, we find
$$
d\mathit \Omega_{0,0}=-d\langle\delta\omega_{0,0}|\hat{H}_+\delta C_{0,2}\rangle
$$
$$
=-\langle D\delta \omega_{0,0}| \hat{H}_+\delta C_{0,2}\rangle-\langle \delta \omega_{0,0}| \hat{H}_+ D\delta C_{0,2}\rangle
$$
$$
\approx \langle \hat{H}_-\delta C_{2,0}+\hat{H}_+\delta C_{0,2}|\hat{H}_+\delta C_{0,2}\rangle-\langle \delta\omega_{0,0}|\hat{H}_+\hat{h}_0\delta C_{1,3}\rangle
$$
$$
=\langle \hat{H}_-\delta C_{2,0}|\hat{H}_+\delta C_{0,2}\rangle=0\,.
$$
The last term vanishes due to the identity $H^{\alpha\beta}_+\wedge H^{\dot\alpha\dot \beta}_-=0$.

From the general considerations of Sec. \ref{PrM} we know that the forms $\varOmega_{mn}$ are  gauge invariant modulo trivial ones. Furthermore, this gauge invariance is a consequence of a more general property that the gauge variations of fields belong to the kernel of the presymplectic form. The last fact can be verified directly. In terms of homogeneous components the gauge symmetry transformations (\ref{GSym}) read
\begin{equation}\label{GSymCom}
\delta_\zeta\omega_{mn}=D\zeta_{mn}+\hat{h}_+\zeta_{m-1,n+1}+\hat{h}_-\zeta_{m+1,n-1}\,,\qquad \delta_\zeta C_{mn}=0\,.
\end{equation}
The right hand sides of these equations define a variational vector field $V$ on the space of fields $\Phi$. Because of the equivalence relation (\ref{ER}) the desired equalities $i_V\Omega_{mn}\simeq 0$ follow immediately  from
$$
i_V\mathit \Omega_{0,n}=-\langle \delta_\zeta\omega_{0,n}|\hat{H}_+\delta C_{0,n+2}\rangle
=-\langle D\zeta_{0,n}+\hat{h}_-\zeta_{1,n-1}|\hat{H}_+\delta C_{0,n+2}\rangle
$$
$$\simeq \langle \zeta_{0,n}|\hat{H}_+ D\delta  C_{0,n+2}\rangle +\langle \zeta_{1,n-1}|\hat{h}_+\hat{H}_+\delta C_{0,n+2}\rangle
$$
$$
\approx \langle \zeta_{0,n}|\hat{H}_+\hat{h}_0\delta C_{1,n+3}\rangle=0\,.
$$

For integer spins it is convenient to represent the equivalence classes (\ref{ER}) by the diagonal elements of the matrix $\hat{\varOmega}$. We set
\begin{equation}\label{osi}
\mathit\Omega_s=\mathrm{Im}\, \Omega_{s-1,s-1}\,,\qquad s=1,2,\ldots\,.
\end{equation}
By definition, the sequence $\{\mathit{\Omega}_s\}_{s=1}^\infty$ consists of real presymplectic currents that are on-shell equivalent to the diagonal elements of the matrix $\hat{\varOmega}$. Notice that the form $\varOmega_s$ depends on the fields of spin $s$.

In particular, for spin-one fields Eq. (\ref{osi}) yields
$$
\mathit{\Omega}_1=-\frac1{2i}\langle\delta\omega_{0,0}|\hat{H}_+\delta C_{0,2}-\hat{H}_-\delta C_{2,0}\rangle\,.
$$
Being rewritten in the vector notation, the last expression is proportional to the presymplectic current for electromagnetic field (\ref{OA}).

For spin-two fields we get
$$
\mathit{\Omega}_2=\frac1{2i}\langle\delta\omega _{1,1}| \hat{h}_+\delta\omega_{0,2}-\hat{h}_-\delta\omega_{2,0}\rangle\,.
$$
Again, one can easily verify that, up to an overall factor, this expression defines nothing but the presymplectic current for the linearized gravity (\ref{s2}).

For half-integer spins it is convenient to choose the representatives of presymplectic currents on the super-diagonal of the matrix $\hat{\varOmega}$, namely,
\begin{equation}\label{oshi}
\mathit{\Omega}_s=\mathrm{Im}\,\Omega_{s-1/2,s-3/2}\,,\quad s=3/2,5/2,\ldots \,.
\end{equation}

The one-parameter family (\ref{osi}), (\ref{oshi}) provides nontrivial presymplectic structures for massless fields of all but two spins. The presymplectic structures for the remaining two cases, $s=0, 1/2$, can be read off from (\ref{osc}) and (\ref{osp}). By making use of the identifications of Sec. \ref{HSFF},  we set
\begin{equation}\label{O01}
\varOmega_0=\langle \delta C_{0,0}|\hat{H}_0 \delta C_{1,1}\rangle\,,\qquad \varOmega_{1/2}=\langle \delta C_{1,0}|\mathcal{\hat{H}}_+\delta C_{0,1} \rangle\,,
\end{equation}
where
$$
\hat{H}_0=h^{\alpha\dot \beta}\wedge h_{\beta\dot\beta}\wedge h^{\beta\dot\alpha}\partial_\alpha\partial_{\dot\alpha}\,,\qquad \mathcal{\hat{H}}_{+}=y_{\alpha}h^{\alpha\dot \beta}\wedge h_{\beta\dot\beta}\wedge h^{\beta\dot\alpha}\partial_{\dot\alpha}\,.
$$
One can also include these presymplectic currents into two families of on-shell closed $(2,3)$-forms, namely,
$$
\varOmega'_{nm}=\langle \delta C_{nm}|\hat{H}_0\delta C_{n+1,m+1}\rangle\,,\qquad \Omega_{nm}''=\langle\delta C_{n+1,m}|\mathcal{\hat{H}}_+\delta C_{n,m+1}\rangle\,.
$$
All these currents, however, appear to be trivial except when  $n=m=0$.

Now the general presymplectic structure on the space of free HS fields can be written as
\begin{equation}\label{PresymStr}
\Omega=\sum_{s=0}^\infty a_s\int_{\Sigma} \varOmega_s\,,\qquad a_s\in \mathbb{R}\,.
\end{equation}
Up to rescalings  of fields of definite spin, $\omega_{^{(s)}}\rightarrow\alpha_s\omega_{^{(s)}}$ and $C_{^{(s)}}\rightarrow \alpha_s C_{^{(s)}}$, this presymplectic structure is equivalent to one defined by the current
$$
\varOmega=\mathrm{Im} \langle\delta \omega|\hat{h}_+\delta\omega - \hat{H}_+\delta C_-\rangle +\mathrm{Im} \langle\delta C|\hat{\mathcal{H}}_+\delta C\rangle +\langle\delta C|\hat{H}_0\delta C\rangle\,.
$$
The anti-de Sitter invariance and nontriviality of the above presymplectic structure  will be proved in the next section.

\section{Symmetries and conservation laws}\label{SCL}

As an immediate application of the covariant presymplectic structure above we are going to derive the conservation laws associated with the anti-de Sitter invariance of the HS equations (\ref{HSE}).
Since the equations  are non-Lagrangian, the first  Noether's theorem \cite{KS} is not directly applicable to them and the presence of global symmetries does not automatically imply the existence of the corresponding conservation laws. Actually, it is the presymplectic structure, rather than the Lagrangian itself, which enables one to convert global symmetries to conservation laws\footnote{The correspondence is not generally one-to-one due to the possible degeneracy of the presymplectic structure for non-Lagrangian dynamics. In the backward direction, it is the concept of Lagrange structure \cite{KazLS} which allows one to relate conservation laws to global symmetries for not necessarily Lagrangian theories \cite{KLS1}, \cite{KLS2}.}. Here by a conservation law we understand a $3$-form $J$ built from the (derivatives of) dynamical fields such that
\begin{equation}\label{dJ}
dJ\approx 0\,.
\end{equation}
Two conservation laws $J$ and $J'$ are considered equivalent if they differ on shell by an exact form, that is, $J\simeq J'$.  Due to the Stoks theorem, equivalent conservation laws share the same charge $Q$ defined by the integral
$$
 Q=\int_\Sigma J\,.
$$
By virtue of (\ref{dJ}) the value of $Q$ is independent of the choice of the Cauchy surface $\Sigma$.

The global symmetries of free HS fields are known to form an infinite-dimensional Lie group. Below  we restrict our consideration to the subgroup of symmetries associated with the isometries of $AdS_4$. These have a clear geometric interpretation and constitute a finite-dimensional subgroup, the anti-de Sitter group $SO(3,2)$. The infinitesimal action of the anti-de Sitter group in the space of free HS fields is given by the relations
\begin{equation}\label{GlSym}
\delta_\xi\omega = [\xi,\omega]_\star+2\hat{H}{}^{_\xi}_+C_-+ 2\hat{H}^{_\xi}_-C_+\,,\qquad
\delta_\xi C=[\xi'',C]_\star+\{\xi', C\}_\star\,,
\end{equation}
where
\begin{equation}\label{xi}
\xi=\xi'+\xi''\,,
\end{equation}
$$
\xi'=-\frac i2\xi^{\alpha\dot\alpha}y_\alpha \bar y_{\dot\alpha}\in \mathcal{F}^0_{1,1}  \,,\qquad \xi''=-\frac i4\xi^{\alpha\beta}y_{\alpha}y_{\beta}-\frac i4\xi^{\dot\alpha\dot\beta}\bar y_{\dot\alpha}\bar y_{\dot\beta}\in \mathcal{F}^0_{2,0}\oplus \mathcal{F}^0_{0,2}
$$
are infinitesimal parameters  and
$$
\hat{H}{}^{_\xi}_-=\xi_{\alpha }{}^{\dot\alpha} h^{\alpha\dot\beta}\partial_{\dot\alpha}\partial_{\dot\beta}\,,\qquad \hat{H}{}^{_\xi}_+=\xi^{\alpha }{}_{\dot\alpha} h^{\beta\dot\alpha}\partial_{\alpha}\partial_{\beta}\,.
$$
The parameters (\ref{xi}) are assumed  to satisfy the condition
\begin{equation}\label{dxi}
\mathcal{D}\xi=D\xi+[h,\xi]=0\,.
\end{equation}
For the derivation and explanation of these formulas  we refer the reader to  \cite[Sec. 5]{V1}, \cite{V2}.

Strange as it may seem, the symmetry transformations (\ref{GlSym}) induced by the isometries of $AdS_4$ do not involve the space-time derivatives of fields.  One important point to remember is that the generators of global symmetries are defined only modulo equations of motion and the unfolded form of dynamics enables the derivatives to be expressed through nondifferentiated fields. Notice also that the symmetry transformations (\ref{GlSym}) do not mix the fields of different spins.

Considering that $\mathcal{D}^2=0$, Eq. (\ref{dxi}) is integrable and allows one  to reconstruct the function $\xi(x)$ by its value $\xi(x_0)$ at any given point $x_0$ of $AdS_4$. We are led  to conclude that there are exactly ten  linearly independent solutions to Eq. (\ref{dxi}). These solutions -- the spin-tensors $\xi$'s -- are naturally identified with the  Killing vectors of the anti-de Sitter metric. Since the operator $\mathcal{D}$ differentiates the $\star$-product,  the solutions to Eq. (\ref{dxi}) form a closed Lie algebra with respect to the $\star$-commutator, the anti-de Sitter algebra. If $\xi_1$ and $\xi_2$ are two such solutions, then $[\delta_{\xi_1},\delta_{\xi_2}]=\delta_{[\xi_1,\xi_2]}$.

By definition, the generators of symmetry transformations (\ref{GlSym}) are given by the variational vector fields that are tangent to the subspace $M\subset \Phi$ of solutions to the HS equations (\ref{HSE}). Let us show that these vector fields are Hamiltonian relative to the presymplectic structure (\ref{PresymStr}). This obviously the case for the fields of low spins ($s\leq 2$) where the presymplectic structures come from the standard Lagrangians.  By the Noether theorem the global symmetries of a Lagrangian give rise to the conserved currents. These currents are nothing but the Hamiltonians that generate the symmetry transformations through the Poisson bracket on the covariant phase space. So, without loss in generality, we can restrict ourselves to the case $s>2$. For higher spins the expressions for  the presymplectic structure (\ref{PresymStr}) and the symmetry transformations (\ref{GlSym})  are considerably simplified.  In terms of homogeneous components, the transformations of the gauge field can be written as
$$
\delta_\xi \omega_{mn} =\hat{\xi}''\omega_{mn} +\hat{\xi}'_{-}\omega_{m+1,n-1}+\hat{\xi}'_+\omega_{m-1,n+1}\,,\qquad mn>0\,,
$$
where
$$
\hat{\xi}''=[\xi'',\,\cdot\,]_\star = \xi^{\alpha\beta}y_\alpha\partial_\beta+\bar{\xi}^{\dot\alpha\dot\beta}\bar{y}_{\dot\alpha}\partial_{\dot\beta}\,,
$$
$$
\hat{\xi}'=[\xi', \,\cdot\,]_\star=\hat{\xi}'_-+\hat{\xi}'_+\,,\qquad \hat{\xi}'_-=\xi^{\alpha\dot\alpha}\bar{y}_{\dot\alpha}\partial_{\alpha}\,,\qquad \hat{\xi}'_+=\xi^{\alpha\dot\alpha}y_{\alpha}\partial_{\dot\alpha}\,.
$$
Let us mention the following useful properties of the introduced operators:
$$
\langle \hat{\xi}'' A|B \rangle=-\langle A|\hat{\xi}'' B\rangle\,,\qquad \langle \hat{\xi}'_+ A|B \rangle=-\langle A|\hat{\xi}'_- B\rangle \qquad \forall A,B\in \mathcal{F}\,,
$$
$$
[\hat{\xi}'_\pm, \hat{h}_{\pm}]=0\,,\qquad [D,\hat{\xi}'_\pm]+[\hat{h}_\pm,\hat{\xi}'']=0\,.
$$

Denoting by $U$ the variational vector field defined by the r.h.s. of equations (\ref{GlSym}) and using  the relations above, one can check that
\begin{equation}\label{BRel}
i_U \varOmega_s\approx \delta J_s + d\Xi_s\,, \qquad s>2\,,
\end{equation}
where
\begin{equation}\label{J}
\begin{array}{rcl}
J_s&=&-\mathrm{Im}\langle \omega_{mn}|\hat{\xi}''\hat{h}_+\omega_{m-1,n+1} +\hat{h}_+\hat{\xi}'_-\omega_{mn}\rangle\\[3mm]
&&-\mathrm{Im}\langle \omega_{m+1,n+1}|\hat{\xi}'_+\hat{h}_+\omega_{m-1,n+1}\rangle
\end{array}
\end{equation}
and
$$
\Xi_s=\mathrm{Im}\langle \delta\omega_{mn}|\hat{\xi}'_+\omega_{m-1,n+1}\rangle\,.
$$
In these formulas one should put $m=n=s-1$ for integer spins and $m=s-1/2$, $n=s-3/2$ for half-integer spins.

Rel. (\ref{BRel}) implies two things. First, applying $\delta$ to both sides of the relation yields
$$
L_U\varOmega_s\simeq 0\,.
$$
This means the anti-de Sitter invariance of the form $\varOmega_s$. Second, acting by the exterior differential $d$ on the left and right hand sides of  (\ref{BRel}), we get $\delta dJ_s\approx 0$.\footnote{Let us mention the useful identity $i_U d+di_U=0$, which holds for all (even) variational vector fields $U$.}  The form $J_s$ being field-dependent, the last relation is equivalent to
$$
dJ_s\approx 0\,.
$$
In other words, the Hamiltonian $J_s$ is conserved. A direct verification shows that the forms $J_s$ are nontrivial. Thus, the presymplectic structure (\ref{PresymStr}) enables us to relate the anti-de Sitter invariance of the HS equations with the conservation laws. Up to trivial redefinitions the forms $J_s$ are seemed to coincide with some of spin-two currents obtained recently in \cite{SV}. In that paper the authors argued that the forms $J_s$ are  bound to be gauge noninvariant whenever $s\geq 2$. This last fact can also be seen from Rel. (\ref{BRel}). In the previous section we have proved the relation
$$i_V\varOmega_s\approx d\Gamma_s\,,$$
where $V$ is the vector field defining the gauge variations of fields (\ref{GSymCom}) and  $\Gamma_s$ is some local $(1,2)$-form. Contracting (\ref{BRel}) with $V$ yields
$$
L_VJ_s\approx d(i_U\Gamma_s+i_V\Xi_s)\,.
$$
We see that, in the general case, the conserved currents $J_s$ are gauge invariant only modulo  $d$-exact forms.

As a byproduct we have shown nontriviality of the presymplectic structure (\ref{PresymStr}): for if it were trivial, then the conserved currents (\ref{J}) would be trivial as well.

\section{Weak Lagrangians}

Let $\varOmega$ be a presymplectic current compatible with (not necessarily Lagrangian) equations of motion (\ref{E}). As is shown in \cite{BHL, Kh}, such a current defines a \textit{weak Lagrangian}. This is given by a top form $\mathcal{L}$ on $N$ built of the fields $\phi^i$ and their derivatives. Regarding $\mathcal{L}$ as a conventional Lagrangian density, one can define the Euler-Lagrange equations
\begin{equation}\label{EL}
\frac{\delta \mathcal{L}}{\delta\phi^i}=0\,.
\end{equation}
The adjective `weak' means that all solutions to (\ref{E}) also solve  (\ref{EL}). In general, the Euler-Lagrange system (\ref{EL}) is not equivalent to the original system of equations, but only to a subsystem thereof, hence the name.

The form $\mathcal{L}$ is constructed with the help of the cohomological descent method. Assuming that the $\delta$-cohomology is trivial\footnote{This is always the case where the target space of fields is contractible \cite{Anderson}. If the equations of motion are regular, then the differential $\delta$ remains acyclic upon restriction to the shell.} in positive degree, we can write the presymplectic current as
$$
\varOmega=\delta\varTheta
$$
for some presymplectic potential current $\varTheta$. By definition, $\varTheta$ is a hybrid form of type $(1,n-1)$. Applying now the de Rham differential $d$ to both sides of the last equation and taking into account (\ref{PF}), we find
$$
\delta d\varTheta\approx 0\,.
$$
Using the acyclicity of $\delta$ once again, we conclude that there exists a $(0,n-1)$-form $\Lambda$ such that
$$
d\varTheta \approx\delta \Lambda\,.
$$
  For regular equations of motion (\ref{E}), the last weak equality can be replaced by the following one:
$$
d\varTheta=\delta \Lambda+ \sum_{n=0}^N\left( \delta E_{a,\mu(n)}\wedge \nu^{a\mu(n)}-E_{a,\mu(n)} \lambda^{a\mu(n)}\right)\,.
$$
Here
$
E_{a,\mu(n)}=\partial_{\mu_1}\cdots\partial_{\mu_n}E_a
$
are the differential consequences of the field equations (\ref{E}); $\lambda$'s and $\nu$'s are some forms of type $(1,n)$ and $(0,n)$, respectively.   By making use of Leibniz's rule for the differential $\delta$, we can rewrite the last relation as
\begin{equation}\label{dE}
\delta\Big(\Lambda+\sum_n E_{a,\mu(n)} \nu^{a\mu(n)}\Big)=d\varTheta+\sum_n E_{a,\mu(n)}(\lambda^{a\mu(n)}+\delta\nu^{a\mu(n)})\,.
\end{equation}
The weak Lagrangian is defined now by
$$
\mathcal{L}=\Lambda+\sum_n E_{a,\mu(n)} \nu^{a\mu(n)}\approx \Lambda\,.
$$
Integrating the r.h.s. of Eq. (\ref{dE}) by parts, we can bring the variation of $\mathcal{L}$ into the standard form
$$
\delta \mathcal{L}=\sum_{n=0}^{\tilde{N}} E_{a,\mu(n)}M_i^{a\mu(n)}\wedge \delta\phi^i+d\Big(\Theta +\sum_{n=0}^{\tilde{N}} E_{a,\mu(n)}\tilde{\lambda}^{a\mu(n)}\Big)
$$
for some $M$'s and $\tilde{\lambda}$'s. This amounts to the equations
$$
\frac{\delta \mathcal{L}}{\delta\phi^i}=\sum_n E_{a,\mu(n)}M_i^{a\mu(n)}\,,\qquad
d\Big(\varTheta +\sum_n E_{a,\mu(n)}\tilde{\lambda}^{a\mu(n)}\Big)= d\varTheta_{\mathcal{L}}\,,
$$
where $\varTheta_\mathcal{L}$ is the current of the canonical presymplectic potential associated with the Lagrangian density $\mathcal{L}$.
 We see that each solution to the original equations of motion $E_a=0$ obeys also the Euler-Lagrange equations (\ref{EL}) for the weak Lagrangian.  The off-shell acyclicity of $d$ implies that $\varTheta \simeq \varTheta_{\mathcal{L}}$; and hence,
$$
\varOmega\simeq\varOmega_{\mathcal{L}}\,.
$$
In other words, the presymplectic structure associated to the weak Lagrangian and the original presymplectic structure are equivalent.

Let the field equations (\ref{E}) enjoy infinitesimal gauge symmetry transformations $\delta_\zeta \phi = V$, i.e.,
$$
L_V E_a=i_V\delta E_a\approx0\,.
$$
Then contracting the form (\ref{dE}) with the variational vector field $V$, we obtain
\begin{equation}\label{LVL}
L_V \mathcal{L}\simeq 0\,.
\end{equation}
This means the on-shell gauge invariance of the weak Lagrangian.

Let us now apply this general construction to the case at hand. In Sec. \ref{CPS}, we introduced the set  of complex presymplectic currents (\ref{Omn}) compatible with the HS field equations (\ref{EoM}). The currents are $\delta$-exact and can be written in the form  
$$
\varOmega_{mn}=\delta \varTheta_{mn}\,,\qquad \mathit{\Theta}_{mn}=\left\{
                       \begin{array}{ll}
                         \langle\omega_{mn}|\hat{h}_+\delta\omega_{m-1,n+1}\rangle, & \hbox{for $m>0$;} \\[3mm]
                         -\langle \omega_{0,n}|\hat{H}_+\delta C_{0,n+2}\rangle, & \hbox{for $m=0$.}
                       \end{array}
                     \right.
$$
Let us first assume that $m>0$. Then, using identities (\ref{id}), we find
\begin{equation}\label{dT}
\begin{array}{c}
d\varTheta_{mn}=\langle D\omega_{mn}|\hat{h}_+ \delta \omega_{m-1,n+1}\rangle
-\langle\omega_{mn}|\hat{h}_+\delta D\omega_{m-1,n+1}\rangle \\[3mm]
=\langle E^\omega_{mn}| \hat{h}_+\delta \omega_{m-1,n+1}\rangle -\langle \hat{h}_+\omega_{m-1,n+1}|\hat{h}_+\delta\omega_{m-1,n+1}\rangle\\[3mm]
-\langle \omega_{mn}|\hat{h}_+\delta E^\omega_{m-1,n+1}\rangle +\langle \omega_{mn}|\hat{h}_+\delta \hat{h}_-\omega_{mn}\rangle\\[3mm]
=\langle E^\omega_{mn}| \hat{h}_+\delta \omega_{m-1,n+1}\rangle +\langle \hat{h}_+\omega_{m-1,n+1}|\delta \hat{h}_+\omega_{m-1,n+1}\rangle\\[3mm]
-\langle \hat{h}_-\omega_{mn}|\delta E^\omega_{m-1,n+1}\rangle +\langle \hat{h}_-\omega_{mn}|\delta \hat{h}_-\omega_{mn}\rangle\\[3mm]
=\delta L_{mn} +\langle \hat{h}_+E^\omega_{m-1,n+1}| \delta\omega_{mn}\rangle -\langle \hat{h}_- E^\omega_{mn}| \delta \omega_{m-1,n+1}\rangle\,,
\end{array}
\end{equation}
where $E^\omega=0$ is the first equation in (\ref{HSE}) or (\ref{EoM}) and
$$
L_{mn}= \frac12\langle \hat{h}_+ \omega_{m-1,n+1}|\hat{h}_+\omega_{m-1,n+1}\rangle +
\frac12 \langle \hat{h}_-\omega_{mn}|\hat{h}_-\omega_{mn}\rangle -
\langle E^\omega_{m-1,n+1}|\hat{h}_-\omega_{mn}\rangle\,.
$$
By construction, the imaginary part of $L_{mn}$ gives us the two-parameter  family of weak Lagrangians
$$
\mathcal{L}_{mn}=\mathrm{Im}\, L_{mn}\,,\qquad m>0\,.
$$
These depend only on the gauge fields.  Rel. (\ref{dT}) implies the following Euler-Lagrange equations:
\begin{equation}\label{hEhE}
\frac{\delta {\mathcal{L}}_{mn}}{\delta \omega_{pq}}=\delta_{p,m-1}\delta_{q,n+1}\hat{h}_-E^\omega_{mn}-\delta_{p,m}\delta_{q,n} \hat{h}_+E^\omega_{m-1,n+1}=0\,.
\end{equation}
We see that the operators $\hat{h}_\pm$ play the role of integrating multipliers for the HS equations. Since $\hat{h}{}^2_\pm=0$, the operators $\hat{h}_\pm$ are characterized by nonzero kernels. In particular, $\mathrm{im}\,\hat{h}_{\pm}\subset \mathrm{ker}\,\hat{h}_\pm$. As a result the Lagrangian equations (\ref{hEhE}) are weaker than the original system (\ref{EoM}).

For $m=0$ we obtain 
$$
d\varTheta_{0,n}=-\langle D\omega_{0,n}|\hat{H}_+\delta C_{0,n+2}\rangle
-\langle \omega_{0,n}|\hat{H}_+\delta DC_{0,n+2}\rangle
$$
$$
=\delta L_{0,n} - \langle \hat{H}_+E^C_{0,n+2}|\delta \omega_{0,n}\rangle-\langle E^\omega_{0,n}|\hat{H}_+\delta C_{0,n+2}\rangle \,,
$$
where $E^C=0$ denotes the second equation in (\ref{HSE}) or (\ref{EoM}) and $$
L_{0,n}=\langle \omega_{0,n}|\hat{H}_+E^C_{0,n+2}\rangle -\frac12\langle \hat{H}_+ C_{0,n+2}|\hat{H}_+C_{0,n+2}\rangle\,.
$$
Taking imaginary part, we get one more family of weak Lagrangians,
$$
\mathcal{L}_{0,n}= \mathrm{Im}\, L_{0,n}\,,
$$
which involves the Weyl fields. Now the most general weak Lagrangian associated with the real presymplectic currents $\{\mathrm{Im}\varOmega_{mn}\}_{m,n=0}^\infty$ reads
$$
\mathcal{L}=\sum_{n,m=0}^\infty \lambda_{nm}\mathcal{L}_{nm}\,,\qquad \lambda_{nm}\in \mathbb{R}\,.
$$ 
As for the presymplectic currents (\ref{O01}), they give the standard Lagrangians for the massless scalar and spinor fields.

 Notice that the Lagrangians $\mathcal{L}_{nm}$ belonging to one and the same spin are on-shell equivalent to each other modulo total divergence:
$$
\mathcal{L}_{nm}\simeq \Lambda_{s}=-\frac12 \mathrm{Im} \langle \hat{H}_+ C_{0,2s}|\hat{H}_+C_{0,2s}\rangle\,,\qquad n+m=2s-2\,.
$$

Due to Rel. (\ref{LVL}), the local functional 
$$
S=\int_{AdS_4} \mathcal{L}
$$
is gauge invariant when evaluated on the solutions to the HS  equations (\ref{HSE}). There are strong grounds to believe that  functionals of this type may be used for establishing the AdS/CFT correspondence. Usually, the role of functional $S$ is played by the `genuine' action of the bulk fields. If such an action is unavailable or does not exist for a given set of bulk fields, one can try to use some other local functionals that are compatible with fundamental symmetries of the model. In the context of nonlinear HS theories, a concrete proposal for the construction of an appropriate  functional $S$ has been put forward in \cite{V3}.

\section{Conclusion}\label{Sum}

 Let us comment on the obtained results and further perspectives. In Sec. \ref{CPS}, we proposed an anti-de Sitter invariant presymplectic structure which is compatible with the unfolded representation for the free HS fields in $AdS_4$.  Although this presymplectic structure may not exhaust all the possibilities, it looks most natural and reproduces the standard presymplectic structures in the case of  low spins. The classification of all possible presymplectic currents  in the free HS theories, especially for colour multiplets, is an interesting open problem.  In many respects it is similar to the classification of the usual conserved currents. The existence of gauge invariant HS currents \cite{GV} suggests that such extra presymplectic structures  are likely to exist in the sector of Weyl fields. As discussed in Sec. \ref{SCL}, every covariant presymplectic structure establishes a specific correspondence between the global HS symmetries and the conservation laws.

Another open problem is finding a nontrivial presymplectic structure for nonlinear HS theories. The problem can be attacked  in two ways. First, one can proceed perturbatively, adding quadratic vertices to the free field equations. Such vertices are now available in more or less explicit form, see \cite{DS}, \cite{BKST}, \cite{DMV}. Deformation of the free equations of motion  implies a compatible deformation of the free presymplectic structure. It would be interesting to find the latter or identify obstructions to its existence. If we take this approach seriously, then the quantum correlators of fields have to involve the vertices coming from the deformed presymplectic structure in addition to those contained in the classical equations of motion.   This also opens the way for the study of quantum anomalies. Second, one can start from the nonlinear Vasiliev's equations and try to find a compatible presymplectic structure. The problem here is worse than it was with the free fields as the nonlinear system involves a bigger Weyl algebra and more field equations.

One way or the other, we see that the covariant presymplectic structure offers a far more flexible approach to the study and quantization of HS theories than the conventional Lagrangian formalism.

\section*{Acknowledgements} The author would like to thank Evgeny Skvortsov for useful discussions and acknowledges support from the Russian Foundation for Basic Research (Project No. 16-02-00284 A).

\end{document}